\documentclass[sigconf]{acmart}

\newcommand{\bh}{\mathbf{h}}

\newcommand{\bz}{\mathbf{z}}
\newcommand{\ttt}{\texttt}
\newcommand{\ie}{\textit{i.e.}}
\newcommand{\eg}{\textit{e.g.}}

\usepackage{multirow}
\usepackage{graphics}
\usepackage{balance}
\usepackage{bbm}
\usepackage{caption}
\usepackage{subcaption}
\usepackage{algorithm}
\usepackage{algorithmic}

\AtBeginDocument{%
  \providecommand\BibTeX{{%
    \normalfont B\kern-0.5em{\scshape i\kern-0.25em b}\kern-0.8em\TeX}}}

\copyrightyear{2021}
\acmYear{2021}
\setcopyright{acmlicensed}\acmConference[CIKM '21]{Proceedings of the 30th ACM International Conference on Information and Knowledge Management}{November 1--5, 2021}{Virtual Event, QLD, Australia}
\acmBooktitle{Proceedings of the 30th ACM International Conference on Information and Knowledge Management (CIKM '21), November 1--5, 2021, Virtual Event, QLD, Australia}
\acmPrice{15.00}
\acmDOI{10.1145/3459637.3482243}
\acmISBN{978-1-4503-8446-9/21/11}
\begin{document}
\fancyhead{}
\title{Contrastive Learning of User Behavior Sequence for Context-Aware Document Ranking
}


\author{Yutao Zhu$^1$, Jian-Yun Nie$^1$, Zhicheng Dou$^2$, Zhengyi Ma$^2$, Xinyu Zhang$^3$ \\ Pan Du$^1$, Xiaochen Zuo$^2$, and Hao Jiang$^3$}
\affiliation{$^1$ Université de Montréal \city{Montréal} \state{Québec} \country{Canada}}
\affiliation{$^2$ Gaoling School of Artificial Intelligence, Renmin University of China \state{Beijing} \country{China}}
\affiliation{$^3$ Distributed and Parallel Software Lab, Huawei \city{Hangzhou} \state{Zhejiang} \country{China}}
\email{yutao.zhu@umontreal.ca,nie@iro.umontreal.ca,{dou,zymaa,zuoxc}@ruc.edu.cn}
\email{du@youark.com,{zhangxinyu35,jianghao66}@huawei.com}

\renewcommand{\shortauthors}{Zhu, et al.}
\renewcommand{\authors}{Yutao Zhu, Jian-Yun Nie, Zhicheng Dou, Zhengyi Ma, Xinyu Zhang, Pan Du, Xiaochen Zuo, and Hao Jiang}

\begin{abstract}
Context information in search sessions has proven to be useful for capturing user search intent. Existing studies explored user behavior sequences in sessions in different ways to enhance query suggestion or document ranking. However, a user behavior sequence has often been viewed as a definite and exact signal reflecting a user's behavior. In reality, it is highly variable: user's queries for the same intent can vary, and different documents can be clicked. To learn a more robust representation of the user behavior sequence, we propose a method based on contrastive learning, which takes into account the possible variations in user's behavior sequences.  Specifically, we propose three data augmentation strategies to generate similar variants of user behavior sequences and contrast them with other sequences. In so doing, the model is forced to be more robust regarding the possible variations. The optimized sequence representation is incorporated into  document ranking. Experiments on two real query log datasets show that our proposed model outperforms the state-of-the-art methods significantly, which demonstrates the effectiveness of our method for context-aware document ranking.

\end{abstract}




\maketitle

\section{Introduction}
Search engines have evolved from one-shot searches to consecutive search interactions with users~\cite{DBLP:conf/sigir/AgichteinWDB12}. To fulfil complex information needs, users will issue a sequence of queries, examine and interact with some of the results. User historical behavior or interaction history in a session is known to be very useful for understanding the user's information needs and to rank documents~\cite{DBLP:conf/cikm/JonesK08,DBLP:conf/sigir/BennettWCDBBC12,DBLP:conf/cikm/GeDJNW18,DBLP:conf/sigir/ZhouDWXW21}.

Various studies exploited user behavior data for different purposes. For example, by analyzing search logs, researchers found that a user's search history provides useful information for understanding user intent during the search sessions~\cite{DBLP:conf/sigir/BennettWCDBBC12}. To utilize the historical user behavior in document ranking, some early work explored query expansion and learning to rank techniques~\cite{DBLP:conf/sigir/BennettWCDBBC12,DBLP:conf/sigir/CarteretteCHKS16,DBLP:conf/ictir/GyselKR16,DBLP:conf/sigir/ShenTZ05}. More recently, various neural structures have been used to model the user behavior sequence. For example, a recurrent neural network (RNN) 
is proposed to model the historical queries and suggest the next query~\cite{DBLP:conf/cikm/SordoniBVLSN15}. This structure has been extended to model both the historical queries and clicked documents, leading to further improvement on document ranking~\cite{CARS,MNSRF}. Pre-trained language models have also been exploited to encode contextual information from user behavior sequences, and they achieved promising results~\cite{HBA}. 

All these studies tried to learn a prediction or representation model to capture the information hidden in the sequences. However, user behavior sequences have been viewed as definite and exact sequences. That is, an observed sequence is used as a positive sample and any unseen sequence is not used or is viewed as a negative sample. This strict view does not reflect the flexible nature of user's behavior in a session. Indeed, when interacting with a search engine, users do not have a definitive interaction pattern, nor a fixed query for an information need. All these are flexible and change greatly from a user to another, and from a search context to another. Similarly, user's click behaviors are also not definitive: one can click on different documents for the same information need, and can also click on irrelevant documents. The high variation is inherent in the user's interactions with a search engine. This characteristic has not been explicitly addressed in previous studies. One typically relied on a large amount of log data, hoping that strong patterns can emerge, while accidental variations (or noise) can be discarded. This is true to some extent when we have a large amount of log data and when we are only interested in the common patterns shared by users. However, the models strictly relying on the log data cannot fully capture the nuances in user behaviors and cope with the variations. A better approach is to view the data as they are, \ie, they are just samples of possible query formulations and interactions, but much more are not shown in the logs. 

To tackle this problem, in this work, we propose a \textbf{data augmentation} approach to generate possible variations from a search log. More specifically, we use three strategies to mask some terms in a query or document, delete some queries or documents, or reorder the sequence. These strategies reflect some typical variations in user's behavior sequences. The generated behavior sequences can be considered similar to the observed ones. We have, therefore, automatically tagged user behavior sequences in terms of similarity, which are precious for model training. In addition, we can generate more training data from search logs, which has always been a critical issue for research in this area. Based on the augmented data, we utilize \textbf{contrastive learning} to extract what is similar and dissimilar. More specifically, the contrastive model tries to pull the similar sequences (generated variants) closer and to distinguish them from semantically unrelated ones. Compared to the existing approaches based on search logs, we expect that contrastive learning can better cope with the inherent variations and generate more robust models to deal with new behavior sequences.

Contrastive learning is implemented with a pre-trained language model BERT~\cite{BERT} through encoding a sequence and its variants into a contextualized representation with a contrastive loss. The document ranking is then learned by a linear projection on top of the optimized sequence representation. With both the original sequences and corresponding variants modeled in the representation, the final ranking function can not only address the context information thoroughly, but also learn to cope with the inherent variations, hence generating better ranking results during prediction.

We conduct experiments on two large-scale real-world search log datasets (AOL and Tiangong-ST). Experimental results show that our proposed method outperforms the existing methods (including those exploiting search logs) significantly, which demonstrates the effectiveness of our approach.

Our contributions are three-fold:

(1) We design three different data augmentation strategies to construct similar sequences of observed user behavior sequences, which modify the original sequence at term, query/document, and behavior levels.

(2) We propose a self-supervised task with a contrastive learning objective based on the augmented behavior sequences to capture what is hidden behind the sequences and their variants, and to distinguish them from other unrelated sequences.

(3) Experiments on two large-scale real-world search log datasets confirm the effectiveness of our method. This study shows that contrastive learning with automatically augmented search logs is an effective way to alleviate the shortage of log data in IR research.
\section{Related Work}\label{sec:related}
\subsection{Exploiting Historical Log Data}
Context information in sessions has shown to be useful in modeling user intent in search tasks~\cite{DBLP:conf/cikm/JonesK08,DBLP:conf/sigir/BennettWCDBBC12,DBLP:conf/cikm/GeDJNW18,DBLP:conf/sigir/ZhouDW20}. 
Early studies focused on extracting contextual features from users' search activities so as to characterize their search intent. For example, some keywords were extracted from users' historical queries and clicked documents and used to rerank the documents for the current query~\cite{DBLP:conf/sigir/ShenTZ05}. Statistical features and rule-based features were also introduced to quantify or characterize context information~\cite{DBLP:conf/cikm/WhiteBD10,DBLP:conf/sigir/XiangJPSCL10}. However, these methods often rely on manually extracted features or handcrafted rules, which limits their application in different retrieval tasks. 

Later, researchers started to build predictive models for users' search intent or future behavior. For example, a hidden Markov model was employed to model the evolution of users' search intent. Then, both document ranking and query suggestion were conducted based on the predicted user intent~\cite{DBLP:conf/www/CaoJPCL09}. Reinforcement learning has also been applied to model user interactions in search tasks~\cite{DBLP:conf/sigir/GuanZY13,DBLP:conf/sigir/LuoZY14}. Unfortunately, the predefined model space or state transition structure limits the learning of rich user-system interactions.

The development of neural networks generated various solutions for context-aware document ranking. Some researchers proposed a hierarchical neural structure with RNNs to model historical queries and suggest the next query~\cite{DBLP:conf/cikm/SordoniBVLSN15}. This model is further extended with the attention mechanism to better represent sessions and capture user-level search behavior~\cite{DBLP:conf/sigir/ChenCCR18}. Recently, researchers found that jointly learning query suggestion and document ranking can boost the model's performance on both tasks~\cite{MNSRF}. In addition to leveraging historical queries, the historical clicked documents are also reported to be helpful in both query suggestion and document ranking~\cite{CARS}. 

More recently, large-scale pretrained language models, such as BERT~\cite{BERT}, have achieved great performance on many NLP and IR tasks~\cite{DBLP:conf/acl/LiuHCG19,DBLP:conf/emnlp/KhashabiMKSTCH20,DBLP:conf/sigir/KhattabZ20,DBLP:conf/wsdm/MaGZFJC21}. \citet{HBA} proposed to concatenate all historical queries, clicked documents, and unclicked documents as a long sequence and leveraged BERT as an encoder to compute their term-level representations. These representations were further combined with relative position embeddings and human behavior embeddings through another transformer-based structure to get the final representations. The ranking score is computed based on the representation of the special ``[CLS]'' token.

Our framework is also based on BERT, but we use contrastive learning to pretrain the model in a self-supervised manner. Theoretically, this strategy 
better leverages the available training data, which can also be applied to existing methods.
\subsection{Contrastive Learning for IR}
Contrastive learning aims to learn effective representation of data by pulling semantically close neighbors together and pushing apart other non-neighbors~\cite{DBLP:conf/cvpr/HadsellCL06,DBLP:conf/icml/0001I20}. It has been widely applied in computer vision~\cite{DBLP:conf/iccv/ZhuangZY19,CMC,SimCLR} and NLP tasks~\cite{CERT,DBLP:journals/corr/abs-2011-01403,CLEAR,SimCSE} and has proven its high efficiency in leveraging the training data without the need of annotation. What is required in contrastive learning is to identify semantically close neighbors. In visual representation, neighbors are commonly generated by two random transformations of the same image (such as flipping, cropping, rotation, and distortion)~\cite{DBLP:conf/nips/DosovitskiySRB14,SimCLR}. Similarly, in text representation, data augmentation techniques such as word deletion, reordering, and substitution are applied to derive similar texts from a given text sequence~\cite{COCO-LM,CLEAR}. 
Although the principle of contrastive learning is well accepted, the ways to implement it are still under exploration, with the general guiding principles of 
\textit{alignment} and \textit{uniformity}
~\cite{DBLP:conf/icml/0001I20}. 

As for pre-training, \citet{DBLP:conf/iclr/ChangYCYK20} designed several paragraph-level pre-training tasks and the Transformer models can improve over the widely-used BM25~\cite{BM25}. \citet{DBLP:conf/wsdm/MaGZFJC21} constructed a representative word prediction (ROP) task for pre-training BERT. Experimental results showed that the BERT model pre-trained with ROP and masked language model (MLM) tasks achieves great performance on ad-hoc retrieval. Our proposed sequence representation optimization stage can be treated as a pre-training stage because it is trained before document ranking (our main task). However, as we do not use external datasets, we do not categorize our method as a pre-training approach.

In this paper, we propose a contrastive learning objective to optimize the sequence representation for improving the downstream document ranking task. This first attempt opens the door to more future studies on applying contrastive learning to IR.

\begin{figure*}[t!]
    \centering
    \includegraphics[width=.9\linewidth]{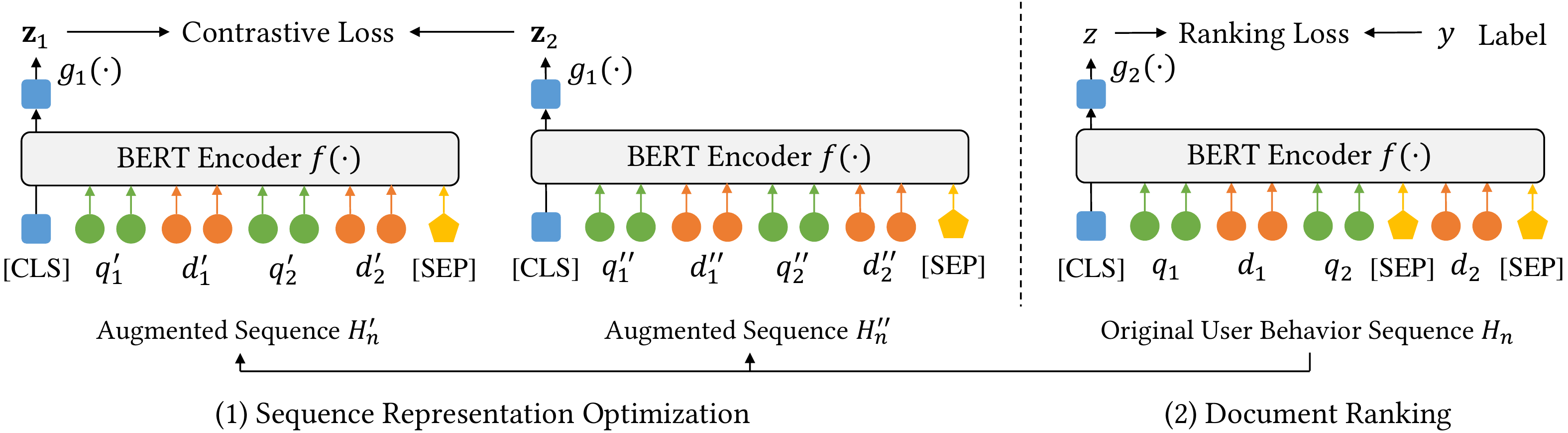}
    \caption{The illustration of \ttt{COCA}. The query-document sequence ($H_n$) is augmented with two different strategies, and the processed sequences are treated as a positive pair ($H'_n$ and $H''_n$). Other augmented sequences in the same minibatch are used to construct negative pairs for them (not shown here). The contrastive objective is to pull close the representation of the two sequences in positive pairs and push apart the representation of others.}
    \label{fig:structure}
    \vspace{-5px}
\end{figure*}

\section{Methodology}\label{sec:method}
Context-aware document ranking aims at using the historical user behavior sequence and the current query to rank a set of candidate documents. In this work, we design a new framework for this task. Our framework aims at optimizing the representation of the user behavior sequence before learning document ranking. As shown in Figure~\ref{fig:structure}, our framework can be divided into two stages: (1) \textit{sequence representation optimization} and (2) \textit{document ranking}. In the first stage, we design a self-supervised task with contrastive learning objective to optimize the sequence representation. In the second stage, our model uses the optimized sequence representation and further learns the ranking model. We call our framework \ttt{COCA} -- \textbf{CO}ntrastive learning for \textbf{C}ontext-\textbf{A}ware document ranking.

\subsection{Notations}
\label{sec:note}
Before introducing the task and the model, we first provide the definitions of important concepts and notations. We present a user's search history as a sequence of $M$ queries $Q=\{q_1,\cdots,q_M\}$, where each query $q_i$ is associated with a submission timestamp $t_i$ and the corresponding list of returned documents $D_i=\{d_{i,1},\cdots,d_{i,M}\}$. Each query $q_i$ is represented as the original text string that users submitted to the search engine. $Q$ is ordered according to query timestamp $t_i$. Each document $d_{i,j}$ has two attributes: its text content and click label $y_{i,j}$ ($y_{i,j}=1$ if it is clicked). In general, user clicks serve as a good proxy of relevance feedback~\cite{DBLP:conf/sigir/JoachimsGPHG05,DBLP:journals/tois/JoachimsGPHRG07,CARS,HBA}. Given all available historical queries and clicked documents up to $n$ turns, we denote the user behavior sequence as $H_n=\{q_1,d_{1},\cdots,q_{n},d_{n}\}$\footnote{Following previous studies~\cite{HBA}, we only use one clicked document to construct the sequence.}. As reported in~\cite{CARS,HBA}, the unclicked documents are less helpful and may even introduce noise, so they are not considered in the user behavior sequence. 



\subsection{Overview}
With the above concepts and notations, we briefly introduce the two stages in \ttt{COCA} as follows.

\noindent (1) \textbf{Sequence Representation Optimization.} As shown in the left side of Figure~\ref{fig:structure}, our target is to obtain a better representation of the user behavior sequence $H_n$ in this stage. To achieve this, we first construct two augmented sequences $H'_n$ and $H''_n$ from $H_n$ with randomly selected augmentation strategies (Section~\ref{sec:aug}). Such a pair of sequences are considered to be similar. Then a BERT encoder is applied to get the representations of these two sequences (Section~\ref{sec:cl_rep}). With the contrastive loss, the model learns to pull them close and push them away from other sequences in the same minibatch (Section~\ref{sec:cl_loss}). By comparing the two augmented sequences, the BERT encoder is forced to learn a more generalized and robust representation for sequences.

\noindent (2) \textbf{Document Ranking.} As shown in the right side of Figure~\ref{fig:structure}, we aim to rank the relevant documents as high as possible in this stage. Given the current query $q_i$ and the historical behavior sequence $H_{i-1}$, we treat $H_{i-1}\cup \{q_i\}$ as a sequence and the candidate document $d_{i,j}$ as another sequence. Then, we concatenate them together and use the BERT encoder trained in the first stage to generate a representation. The final ranking score is obtained by a linear projection on the representation. A cross-entropy loss is applied between the predicted ranking score and the click label $y_{i,j}$.


\subsection{Sequence Representation Optimization}
The user behavior sequence contains abundant information about the user intent. 
To optimize the representation of the user behavior sequence, we propose a self-supervised approach. 
Specifically, we apply a contrastive learning objective to pull close the representation of similar sequences and push apart different ones. The similar sequences are created by the three augmentation strategies described below.

\begin{figure}[t!]
    \centering
    \includegraphics[width=.95\linewidth]{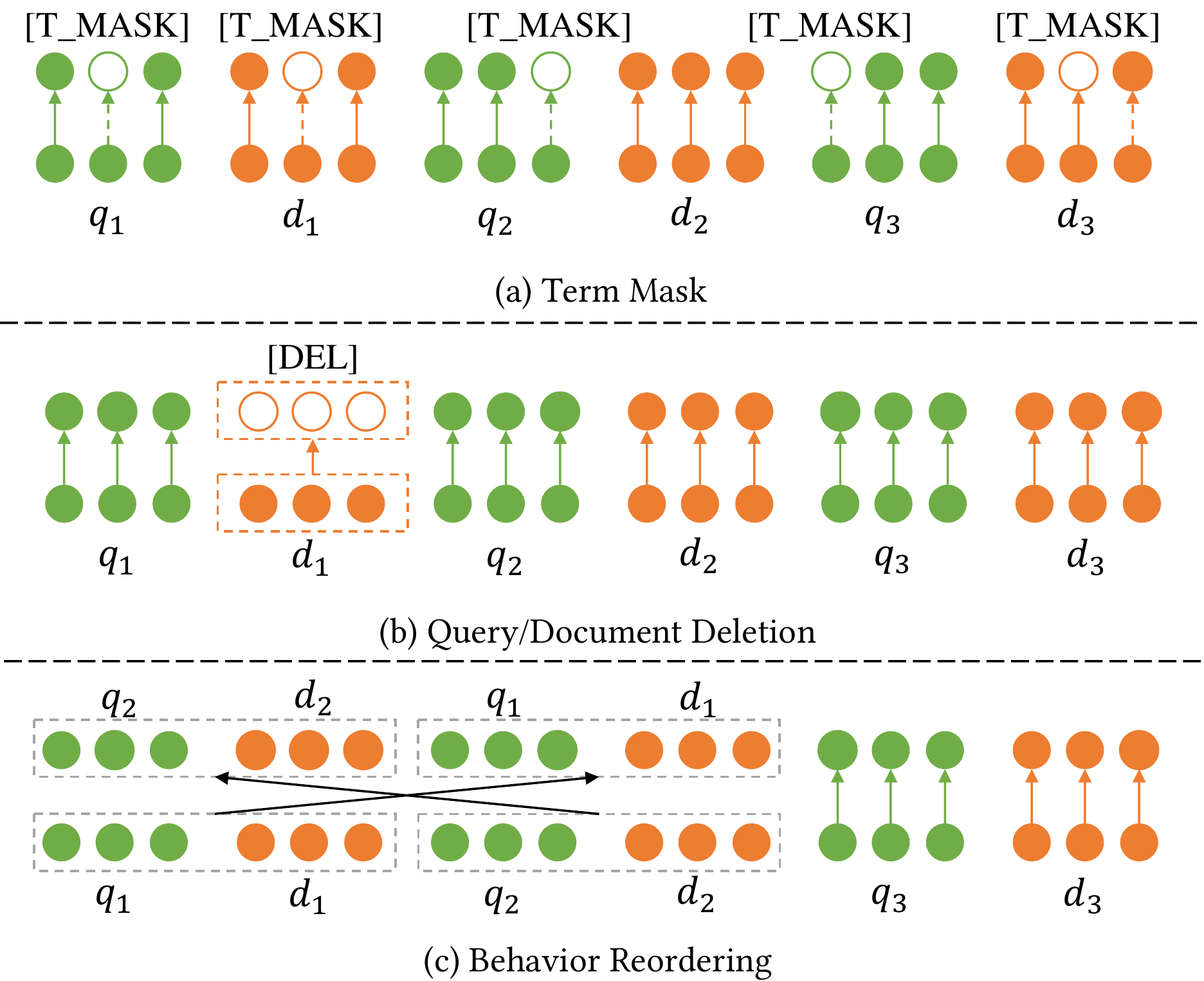}
    \caption{Three augmentation strategies used in \ttt{COCA}. We use the user behavior sequence with three query-document pairs as an example.}
    \label{fig:augmentation}
\end{figure}

\subsubsection{Augmentation Strategy}\label{sec:aug}
Inspired by the existing data augmentation strategies in NLP and image processing, we propose three strategies to construct similar sequences, namely term mask, query/document deletion, and behavior reordering (shown in Figure~\ref{fig:augmentation}). These strategies correspond to three levels of variation in user behaviors, \ie, term level, query/document level, and user behavior level.

\noindent (a) \textbf{Term Mask.} In natural language processing, the ``word mask'' or ``word dropout'' technique has been widely applied to avoid overfitting. It has been shown to improve the robustness of the sentence representation, \eg, in sentence generation~\cite{DBLP:conf/conll/BowmanVVDJB16}, sentiment analysis~\cite{DBLP:conf/nips/DaiL15}, and question answering~\cite{DBLP:conf/comad/GNG18}. Inspired by this, we propose to apply a random term mask operation over the user behavior sequence (including query terms and document terms) as one of the augmentation strategies for contrastive learning. 

With the term-level augmentation strategy, we can obtain various user behavior sequences similar to the original one. The similar sequences only have minor differences in some terms. This aims to simulate the real search situations where users may issue slightly different queries for searching the same target, and a document may satisfy similar information needs. By contrasting similar sequences with others, the models can learn the importance of different terms in both queries and documents. Besides, it can also help the model to learn more generalized sequence representation by avoiding relying too much on specific terms.

Specifically, for a user behavior sequence $H_n=\{q_1,d_1,\cdots,q_n,d_n\}$, we first represent it as a term sequence $H_n=\{w_1,\cdots,w_{N_T}\}$, where $N_T$ is the total number of terms. Then, we randomly mask a proportion $\gamma$ of terms $T_n=\{t_1,\cdots,t_L\}$, where $L_{\rm tm}=\lfloor N\cdot\gamma\rfloor$, and $t_i$ is the index of term to be masked. If a term is masked, it is replaced by a special token ``[T\_MASK]'', which is similar to the token ``[MASK]'' used in BERT~\cite{BERT}. Therefore, we formulate this augmentation strategy as a function $f^{\rm tm}$ over the user behavior sequence $H_n$ as:
\begin{align}
    f^{\rm tm}(H_n) &= \{\widehat{w}_1,\cdots,\widehat{w}_N\}, \\
    \widehat{w}_t &= 
    \begin{cases}
        w_t, & t \notin T_n, \\
        {\rm [T\_MASK]}, & t \in T_n.
    \end{cases}
\end{align}

\noindent (b) \textbf{Query/Document Deletion.} Random crop (deletion) is a common data augmentation strategy in computer vision to increase the variety of images~\cite{SimCLR,CMC}. This operation can create a random subset of an original image and help the model generalize better. Inspired by this, we propose a query/document deletion augmentation operation for contrastive learning. 

The query/document deletion strategy can improve the learning of sequence representation in two respects. First, after deletion, the resulting 
user behavior sequence becomes a similar one with the difference on some queries or documents. This reflects a type of variation in real query logs. By contrasting these similar sequences with others, the models are trained to learn the influence of the deleted queries or documents. Second, the generated incomplete sequence provides a partial view of the original sequence, which forces the model to learn a more robust representation without relying on complete information.

Specifically, for a user behavior sequence $H_n=\{q_1,d_1,\cdots,q_n,d_n\}$, we treat each query and document as a sub-sequence $s$ and represent the sequence as $H_n=\{s_1,s_2,\cdots,s_{2n-1},s_{2n}\}$. Then, we randomly delete a proportion $\mu$ of sub-sequences $R_n=\{r_1,\cdots,r_L\}$, where $L_{\rm del}=\lfloor2n\cdot\mu\rfloor$, and $r_i$ is the index of the sub-sequence to be deleted. Different from the term mask strategy, if a query or document is deleted, the whole sub-sequence is replaced by a special token ``[DEL]''. This augmentation strategy is formulated as a function $f^{\rm del}$ on $H_n$ and defined as:
\begin{align}
    f^{\rm del}(H_n) &= \{\widehat{s}_1,\cdots,\widehat{s}_{2n}\}, \\
    \widehat{s}_r &= 
    \begin{cases}
        s_r, & r \notin R_n, \\
        {\rm [DEL]}, & r \in R_n.
    \end{cases}
\end{align}

\noindent (c) \textbf{Behavior Reordering.} Many tasks assume the strict order of the sequence, \eg, natural language generation~\cite{DBLP:conf/emnlp/KikuchiNSTO16,Transformer} and text coherence modeling~\cite{DBLP:conf/emnlp/LiH14a,DBLP:conf/emnlp/LiJ17,DBLP:conf/aaai/ZhuZNLD21}. However, we observe that the user search behavior sequence is much more flexible. For example, when users only have a vague search intent, they will issue several queries in a random order to obtain related information before making  their real intent clear~\cite{DBLP:conf/sigir/GuanZY13}. Besides, sometimes users may issue a repeated query when they miss some information, which is called re-finding behavior~\cite{DBLP:conf/cikm/MaDBW20,DBLP:conf/wsdm/ZhouDW20}. Under this circumstance, we cannot assume the order of the queries is strict. To prevent the model from relying too much on the order information and make the model more robust to the newly issued query, we propose a behavior reordering strategy for contrastive learning. Different from the former two strategies, user behavior reordering does not reduce the information contained in the sequence. Models can focus on learning content representation in queries and documents rather than merely ``remembering'' their relative order.

For a user behavior sequence $H_n=\{q_1,d_1,\cdots,q_n,d_n\}$, we treat each query and its corresponding document as a behavior sub-sequence and denote it as $H_n=\{b_1,\cdots,b_n\}$, where $b_i=\{q_i,d_i\}$. Then, we randomly select two behavior sub-sequences  and switch their positions, and this operation is conducted $L_{\rm br}=\lfloor n\cdot\eta\rfloor$ times, where $\eta$ is the reordering ratio. Considering the randomly selected $i$-th pairwise position as $(u_i, v_i)$, we switch $b_{u_i}$ and $b_{v_i}$, which can be formulated as a function $f^{br}$ on $H_n$:
\begin{align}
    f^{br}(H_n) &= \{\widehat{b}_1,\cdots,\widehat{b}_n\} \\
    \widehat{b}_j &= 
    \begin{cases}
        b_j, & j \neq u_i \;{\rm and}\; j\neq v_i, \\
        b_{v_i}, & j = u_i, \\
        b_{u_i}, & j = v_i.
    \end{cases}
\end{align}

\subsubsection{Representation}
\label{sec:cl_rep}
Previous work has shown the effectiveness of applying BERT~\cite{BERT} for sequence representation~\cite{DBLP:conf/sigir/DaiC19,DBLP:journals/corr/abs-1901-04085,HBA,DBLP:conf/aaai/ZhuZNLD21,DBLP:conf/sigir/SuDZQW21}. In our framework, we also use the pre-trained BERT as an encoder to represent the augmented user behavior sequences (shown in the left side of Figure~\ref{fig:structure}).
For a user behavior sequence $H_n=\{q_1,d_1,\cdots,q_n,d_n\}$, following the design of vanilla BERT, we add special tokens ``[CLS]'' and ``[SEP]'' at the head and tail of the sequence, respectively. Besides, to further indicate the end of each query/document, we append a special token ``[EOS]'' at the end of it. Therefore, the input sequence $X$ is represented as:
\begin{align}
    X = {\rm [CLS]} q_1 {\rm [EOS]} d_1 {\rm [EOS]} \cdots q_n {\rm [EOS]} d_n {\rm [EOS]} {\rm [SEP]}.
\end{align}
Then, the embedding of each token, the positional embedding, and the segment embedding\footnote{Please refer to the original paper of BERT~\cite{BERT} for more details about these embeddings.} are added together and input to BERT to obtain the contextualized representation. The output of BERT is a sequence of representations for all tokens, and we use the representation of ``[CLS]'' token as the sequence representation:
\begin{align}
    \bz = g_1\big({\rm BERT}(X)_{\rm [CLS]}\big),
\end{align}
where $\bz\in\mathbb{R}^{768}$, and $g_1(\cdot)$ is a linear projection.

\subsubsection{Training Objective}
\label{sec:cl_loss}
We apply a contrastive learning objective to optimize the user behavior sequence representation. A contrastive learning loss function is defined for the contrastive prediction task, \ie, trying to predict the positive augmentation pair $(H_i, H_j)$ in set $\{\mathcal{H}\}$. We construct the set $\{\mathcal{H}\}$ by randomly augmenting twice for all sequences in a minibatch. The strategy of each augmentation is randomly selected from our proposed three ones. Assuming a minibatch with size $N$, we can obtain a set $\{\mathcal{H}\}$ with size $2N$. The two augmented sequences from the \textit{same} user behavior sequence form the positive pair, while all other sequences from the same minibatch are regarded as negative samples for them. Following previous work~\cite{SimCLR,SimCSE,CLEAR,CERT}, the contrastive learning loss for a positive pair is defined as:
\begin{align}
\label{eq:cl}
    l(i,j) = -\log\frac{\exp({\rm sim}(\bz_i,\bz_j)/\tau)}{\sum_{k=1}^{2N}\mathbbm{1}_{k\neq i}\exp({\rm sim}(\bz_i,\bz_k)/\tau)},
\end{align}
where $\mathbbm{1}_{k\neq i}$ is the indicator function to judge whether $k\neq i$ and $\tau$ is a hyperparameter representing temperature. The overall contrastive learning loss is defined as all positive pairs' losses in a minibatch:
\begin{align}
    \mathcal{L}_{\rm CL} = \sum_{i=1}^{2N}\sum_{j=1}^{2N}m(i,j)l(i,j),
\end{align}
where $m(i,j)=1$ when $(H_i, H_j)$ is a positive pair, and $m(i,j)=0$ otherwise.

From another perspective, the contrastive learning stage can be viewed as a kind of domain-specific post-training for pre-trained language models. As these contextualized language models are usually pre-trained on general corpora, such as the Toronto Books Corpus and Wikipedia, it is less effective to directly fine-tune these models on our downstream ranking task if there is a domain shift. Our contrastive learning stage can help the model on domain adaptation to further improve the ranking task. This strategy has shown to be effective in various tasks including reading comprehension~\cite{DBLP:conf/naacl/XuLSY19} and dialogue generation~\cite{DBLP:journals/corr/abs-2009-04703,SABERT}. 

\subsection{Context-aware Document Ranking}
In the previous step, the BERT encoder has been optimized with the contrastive learning objective. We now incorporate this BERT encoding to  learn the context-aware document ranking task.
\subsubsection{Representation}
Previous studies have applied BERT for ranking in a manner of sequence pair classification~\cite{DBLP:conf/sigir/DaiC19,DBLP:journals/corr/abs-1901-04085,HBA}. Different from the first stage, the ranking stage aims at measuring the relationship between the historical user behavior sequence $H_{n-1}=\{q_1,d_1,\cdots,q_{n-1},d_{n-1}\}$, the current query $q_n$, and a candidate document $d_{n,i}$. Therefore, we treat $H_{n-1}\cup q_n$ as one sequence and $d_{n,i}$ as another sequence, and the input sequence $Y$ is represented as:
\begin{align}
    Y = {\rm [CLS]}q_1{\rm [EOS]}d_1{\rm [EOS]}\cdots q_n{\rm [EOS]}{\rm [SEP]}d_{n,i}{\rm [EOS]}{\rm [SEP]}. \notag
\end{align}
Afterwards, the embedding of each token, the positional embedding, and the segment embedding are added together and input to BERT. Note that $Y$ contains two sequences, so we set their segment embeddings respectively as 0 and 1 to distinguish them. The output representation of ``[CLS]'' is used as the sequence representation to calculate the ranking score $z$ as:
\begin{align}
    \bh = {\rm BERT}(Y)_{\rm [CLS]}, \quad z = g_2(\bh),
\end{align}
where $\bh\in\mathbb{R}^{768}$, and $g_2(\cdot)$ is a linear projection to map the representation into a (scalar) score.

\subsubsection{Optimization}
Following previous studies~\cite{CARS,HBA}, we use the following cross-entropy loss to optimize the model:
\begin{align}
    \mathcal{L}_{\rm rank} = - \frac{1}{N}\sum_{i=1}^{N} y_i\log z_i + (1-y_i)\log (1-z_i),
\end{align}
where $N$ is the number of samples in the training set.

\section{Experiments}\label{sec:exp}
\subsection{Datasets and Evaluation Metrics}
We conduct experiments on two public datasets: AOL search log data\footnote{We understand that the AOL dataset should normally not be used in experiments. We choose to use it here because it contains real human clicks, which fits our experiments well. MS MARCO Conversational Search dataset may be another possible dataset, but the sessions in it are artificially constructed rather than real search logs. So, we do not use the MS MARCO dataset in experiments.}~\cite{AOL} and Tiangong-ST query log data~\cite{Tiangong}. 

For AOL search log, we use the one provided by~\citet{CARS}. The dataset contains a large number of sessions, and each session consists of several queries. In training and validation sets, there are five candidate documents for each query in the session. In the test set, 
50 documents retrieved by BM25~\cite{BM25} are used as candidates for each query in the session. All queries have at least one satisfied click in this dataset, and if there are more than one clicked documents, we use the first one in the list to construct the user behavior sequence.

Tiangong-ST dataset is collected from a Chinese commercial search engine. It contains web search session data extracted from an 18-day search log. Each query in the dataset has 10 candidate documents. In the training and validation sets, we use the clicked documents as the satisfied clicks. Some queries may have no satisfied click, we use a special token ``[Empty]'' for padding. 
For the test, the last query of each session is manually annotated with relevance scores, while other (previous) queries in the session have only click labels. Therefore, we construct two test sets based on the original test data as follows:

(1) Tiangong-ST-Click: In this test set, we only use the previous queries (\ie, without the last query) and their candidate documents. Similar to AOL dataset, in this test scenario, all documents are labeled with ``click'' or ``unclick'', and the model is asked to rank the clicked documents as high as possible. Note that the query with no click document is not used for testing.

(2) Tiangong-ST-Human: In this test set, only the last query with human annotated relevance score is used. The score ranges from 0 to 4. More details can be found in~\cite{Tiangong}.

The statistics of both datasets are shown in Table~\ref{tab:sta}. Following previous studies~\cite{DBLP:conf/cikm/GaoHN10,DBLP:conf/cikm/HuangHGDAH13,DBLP:conf/ijcai/HuangZSWL18,CARS}, to reduce memory requirements and speed up training, we only use the document title as its content.

\textbf{Evaluation Metrics} We use Mean Average Precision (MAP), Mean Reciprocal Rank (MRR), and Normalized Discounted Cumulative Gain at position $k$ (NDCG@$k$, $k=\{1,3,5,10\}$) as evaluation metrics. For Tiangong-ST-Human, since the candidate documents are provided by a commercial search engine, the irrelevant documents are expected to be  limited. Hence, as suggested by the authors of~\cite{Tiangong}, we only evaluate the results with NDCG@$k$. All evaluation results are calculated by TREC's evaluation tool (trec\_eval)~\cite{DBLP:conf/sigir/GyselR18}.

\begin{table}[t!]
    \centering
    \small
    \caption{The statistics of the datasets used in the paper.}
    \begin{tabular}{lrrrrrr}
    \toprule
        \textbf{AOL} & \textbf{Training} & \textbf{Validation} & \textbf{Test} \\
    \midrule
        \# Sessions & 219,748 & 34,090 & 29,369 \\
        \# Queries  & 566,967 & 88,021 & 76,159 \\
        Avg. \# Query per Session & 2.58 & 2.58 & 2.59 \\
        Avg. \# Document per Query & 5 & 5 & 50 \\
        Avg. Query Len & 2.86 & 2.85 & 2.9 \\
        Avg. Document Len & 7.27 & 7.29 & 7.08 \\   
        Avg. \# Clicks per Query & 1.08 & 1.08 & 1.11 \\
    \midrule
        \textbf{Tiangong-ST} & \textbf{Training} & \textbf{Validation} & \textbf{Test} \\
    \midrule
        \# Sessions & 143,155 & 2,000 & 2,000 \\
        \# Queries  & 344,806 & 5,026 &  6,420 \\
        Avg. \# Query per Session & 2.41 & 2.51 & 3.21 \\
        Avg. \# Document per Query & 10 & 10 & 10 \\
        Avg. Query Len & 2.89 & 1.83 & 3.46 \\
        Avg. Document Len & 8.25 & 6.99 & 9.18 \\
        Avg. \# Clicks per Query & 0.94 & 0.53 & (3.65) \\
    \bottomrule
    \end{tabular}
    \label{tab:sta}
\end{table}

\begin{table*}[t!]
    \centering
    \small
    \caption{Experimental results on all datasets. All baseline models are based on the code released in the original paper. The best performance and the second best performance are in bold and underlined, respectively. The improvement of \ttt{COCA} over the best baseline is given in the last column. $\dag$ indicates \ttt{COCA} achieves significant improvements over all existing methods in paired t-test with $p$-value $<$ 0.01.}
    \begin{tabular}{llcccccccccr}
    \toprule
        Dataset & Metric & \ttt{ARC-I} & \ttt{ARC-II} & \ttt{KNRM} & \ttt{Duet} & \ttt{M-NSRF} & \ttt{M-Match} & \ttt{CARS} & \ttt{HBA} & \ttt{COCA} & Improv. \\
        \midrule
        \multirow{6}{*}{AOL} & MAP & 0.3361 & 0.3834 & 0.4038 & 0.4008 & 0.4217 & 0.4459 & 0.4297 & \underline{0.5281} & \textbf{0.5500}$^\dag$ & 4.15\% \\
        & MRR & 0.3475 & 0.3951 & 0.4133 & 0.4111 & 0.4326 & 0.4572 & 0.4408 & \underline{0.5384} & \textbf{0.5601}$^\dag$ & 4.03\% \\
        & NDCG@1 & 0.1988 & 0.2428 & 0.2397 & 0.2492 & 0.2737 & 0.3020 & 0.2816 & \underline{0.3773} & \textbf{0.4024}$^\dag$ & 6.65\% \\
        & NDCG@3 & 0.3108 & 0.3564 & 0.3868 & 0.3822 & 0.4025 & 0.4301 & 0.4117 & \underline{0.5241} & \textbf{0.5478}$^\dag$ & 4.52\% \\
        & NDCG@5 & 0.3489 & 0.4026 & 0.4322 & 0.4246 & 0.4458 & 0.4697 & 0.4542 & \underline{0.5624} & \textbf{0.5849}$^\dag$ & 4.00\% \\
        & NDCG@10 & 0.3953 & 0.4486 & 0.4761 & 0.4675 & 0.4886 & 0.5103 & 0.4971 & \underline{0.5951} & \textbf{0.6160}$^\dag$ & 3.51\% \\ 
        \midrule
        \multirow{6}{*}{Tiangong-ST-Click} & MAP & 0.6597 & 0.6729 & 0.6551 & 0.6745 & 0.6836 & 0.6778 & {0.6909} & \underline{0.6957} & \textbf{0.7481}$^\dag$ & 7.53\% \\
        & MRR & 0.6826 & 0.6954 & 0.6748 & 0.7026 & 0.7065 & 0.6993 & {0.7134} & \underline{0.7171} & \textbf{0.7696}$^\dag$ & 7.32\% \\
        & NDCG@1 & 0.5315 & 0.5458 & 0.5104 & 0.5738 & 0.5609 & 0.5499 & {0.5677} & \underline{0.5726} & \textbf{0.6386}$^\dag$ & 11.53\%\\
        & NDCG@3 & 0.6383 & 0.6553 & 0.6415 & 0.6511 & 0.6698 & 0.6636 & {0.6764} & \underline{0.6807} & \textbf{0.7445}$^\dag$ & 9.37\% \\
        & NDCG@5 & 0.6946 & 0.7086 & 0.6949 & 0.6955 & 0.7188 & 0.7199 & {0.7271} & \underline{0.7292} & \textbf{0.7858}$^\dag$ & 7.76\% \\
        & NDCG@10 & 0.7509 & 0.7608 & 0.7469 & 0.7621 & 0.7691 & 0.7646 & {0.7746} & \underline{0.7781} & \textbf{0.8180}$^\dag$ & 5.13\% \\
        \midrule
        \multirow{4}{*}{Tiangong-ST-Human} & NDCG@1 & 0.7088 & 0.7131 & 0.7560 & 0.7577 & 0.7124 & 0.7311 & 0.7385 & \underline{0.7612} & \textbf{0.7769} & 2.06\% \\
        & NDCG@3 & 0.7087 & 0.7237 & 0.7457 & 0.7354 & 0.7308 & 0.7233 & 0.7386 & \underline{0.7518} & \textbf{0.7576} & 0.77\% \\
        & NDCG@5 & 0.7317 & 0.7379 & 0.7716 & 0.7548 & 0.7489 & 0.7427 & 0.7512 & \underline{0.7639} & \textbf{0.7703} & 0.84\% \\
        & NDCG@10 & 0.8691 & 0.8732 & 0.8894 & 0.8829 & 0.8795 & 0.8801 & 0.8837 & \underline{0.8896} & \textbf{0.8932} & 0.40\% \\ 
    \bottomrule
    \end{tabular}
    \label{tab:result}
\end{table*}

\subsection{Baseline}
We compare our method with several baseline methods, including those for (1) ad-hoc ranking and (2) context-aware ranking.

(1) \textbf{Ad-hoc ranking methods}. These methods do not use context information (historical queries and documents), and only current query is used for ranking documents. 
\ttt{KNRM}~\cite{KNRM} performs fine-grained interaction between current query and candidate documents and obtain a matching matrix. The ranking features and scores are then calculated by a kernel pooling method.
\ttt{ARC-I}~\cite{ARC-I} is a representation-based method. The query and document are represented by convolutional neural networks (CNNs), respectively. The score is calculated by a multi-layer perceptron (MLP).
\ttt{ARC-II}~\cite{ARC-I} is an interaction-based method. A matching map is constructed from the query and document, based on which the matching features are extracted by CNNs. The score is also computed by an MLP.
\ttt{Duet}~\cite{DUET} computes local and distributed representations of the query and document by several layers of CNNs and MLPs. Then, it integrates both interaction-based features and representation-based features to compute ranking scores. 

(2) \textbf{Context-aware ranking methods}. These methods can leverage both context information and current query to rank candidate documents. 
\ttt{M-NSRF}~\cite{MNSRF} is a multi-task model, which jointly predicts the next query and ranks corresponding documents. The historical queries in a session are encoded by a recurrent neural network (RNN). The ranking score is computed based on the query representation, history representation, and document representation.
\ttt{M-Match-Tensor}~\cite{MNSRF} is similar to \ttt{M-NSRF} but learns a contextual representation for each word in the queries and documents. The computation of ranking score is based on the word-level representation.
\ttt{CARS}~\cite{CARS} is also a multi-task model, which learns query suggestion and document ranking simultaneously. Different from \ttt{M-NSRF}, this method also models the click documents in the history through an RNN. An attention mechanism is applied to compute representations for each query and document. The final ranking score is computed based on the representation of historical queries, clicked documents, current query, and candidate documents\footnote{We will notice some slight discrepancies between our results and those of the original paper of CARS. This is due to different tie-breaking strategies in evaluation. Following~\cite{HBA}, we use trec\_eval while the authors of CARS uses an author-implemented evaluation.}.
\ttt{HBA-Transformer}~\cite{HBA} (henceforth denoted as \ttt{HBA}) concatenates historical queries, clicked documents, and unclick documents into a long sequence and applies BERT~\cite{BERT} to encode them into representations. Then, a higher-level transformer structure with behavior embedding and relative position embedding is employed to further enhance the representation. Finally, the representation of the first token (``[CLS]'') is used to calculate the ranking score. This is the state-of-the-art method in context-aware document ranking task. It is the most similar to our approach, but without contrastive learning.

\subsection{Implement Details}
We use PyTorch~\cite{DBLP:conf/nips/PaszkeGMLBCKLGA19} and Transformers~\cite{DBLP:journals/corr/abs-1910-03771} to implement our model. The pre-trained BERT is provided by Huggingface\footnote{\url{https://huggingface.co/bert-base-uncased}}. The maximum number of tokens in the two stages are set as 128. Sequences with more than 128 tokens are truncated by popping query-document pairs from the head. We use AdamW~\cite{DBLP:conf/iclr/LoshchilovH19} optimizer in both stages. In the sequence representation optimization stage, both the term mask ratio and query/document deletion ratio are tuned from 0.1 to 0.9 and set as 0.6. As for behavior reordering, only one pair of positions are switched because the session is not long (on average 2.5 queries per sessions). The three strategies are randomly selected. Note that the reordering strategy can only be applied to sessions with more than one query. The batch size is set as 128, and the temperature is set as 0.1. We train the model for four epochs. The learning rate is set as 5e-5. In the document ranking stage, we apply a dropout layer on the sequence representation with the rate of 0.1. The learning rate is set as 5e-5 and linearly decayed during the training. We train the model for three epochs. All hyperparameters are tuned based on the performance on the validation set. Our code is released on GitHub at \url{https://github.com/DaoD/COCA}.

\subsection{Experimental Results and Analysis}
The experimental results are shown in Table~\ref{tab:result}. We can find \ttt{COCA} outperforms all existing methods. This result clearly demonstrates the superiority of our method. Based on the results, we can make the following observations.

(1) Among all models, \ttt{COCA} achieves the best results, which demonstrates its effectiveness on modeling user behavior sequence through contrastive learning. In general, the context-aware document ranking models perform better than ad-hoc ranking models. For example, on AOL dataset, the weak contextualized model \ttt{M-NSRF} can still outperform the strong ad-hoc ranking model \ttt{KNRM}. This indicates that modeling user behavior sequence is beneficial for understanding user intent and improving the ranking results. 

(2) Compared with the RNN-based multi-task learning models (\ttt{M-NSRF}, \ttt{M-Match-Tensor}, and \ttt{CARS}), BERT-based methods (\ttt{HBA} and \ttt{COCA}) achieve better performance. Specifically, on AOL dataset, \ttt{HBA} and \ttt{COCA} improve the results by more than 15\% in terms of all metrics. It is worth noting that \ttt{HBA} and \ttt{COCA} learn document ranking independently without supervision signals from query suggestion task. This result reflects the clear advantage of applying pre-trained language models (\eg, BERT) in document ranking.

(3) \ttt{HBA} is the state-of-the-art method on context-aware document ranking task. It designs complex structures over a BERT encoder to consider user behavior in various aspects, including an intra-behavior attention on clicked documents and skipped documents; an inter-behavior attention on all turns; and an embedding indicates their relative positions. In contrast, our \ttt{COCA} only applies a standard BERT encoder and achieves significantly better performance (paired t-test with $p$-value $<$ 0.01). Both MAP and MRR are improved about 4\%. The key difference between them is the contrastive learning we use. The improvements of \ttt{COCA} over \ttt{HBA} directly reflects the advantage of using contrastive learning for behavior sequence representation. 

(4) Intriguingly, the improvements of \ttt{COCA} on AOL and Tiangong-ST-Click are much more significant than that on Tiangong-ST-Human test set. The potential reasons fall in two aspects: 
(a) \ttt{COCA} is trained on data with click labels rather than relevance labels, and the construction of the user behavior sequence is also based on click labels. Therefore, the model is better at predicting click-based scores than relevance scores.
(b) According to our statistics, there are more than 77.4\% documents labeled as relevant (\ie, their annotated relevance scores are larger than 1), so the base score is very high. Even the basic model \ttt{ARC-I} can achieve 0.7088 and 0.8691 in terms of NDCG@1 and NDCG@10. Without more accurate relevance labels for training, it is more difficult for our model to further improve relevance ranking.

\subsection{Discussion}
We further investigate the following research questions.

\begin{table}[t!]
    \centering
    \small
    \caption{Performance of \ttt{COCA} on the AOL dataset with different data augmentation strategies.}
    \setlength{\tabcolsep}{1.5mm}{
    \begin{tabular}{lccccc}
    \toprule
        & MAP & MRR & NDCG@1 & NDCG@3 & NDCG@10 \\
    \midrule
        \ttt{COCA (Full)} & \textbf{0.5500} & \textbf{0.5601} & \textbf{0.4024} & \textbf{0.5478} & \textbf{0.6160} \\
        \quad \ttt{None} & 0.5341 & 0.5445 & 0.3867 & 0.5296 & 0.5999 \\
        \quad \ttt{TM} & 0.5472 & 0.5576 & 0.4009 & 0.5444 & 0.6121 \\
        \quad \ttt{QDD} & 0.5452 & 0.5554 & 0.3969 & 0.5422 & 0.6110 \\
        \quad \ttt{TM + QDD} & 0.5492 & 0.5592 & 0.4005 & 0.5467 & 0.6155 \\
        \quad \ttt{TM + BR} & 0.5448 & 0.5550 & 0.3963 & 0.5414 & 0.6115 \\
        \quad \ttt{QDD + BR} & 0.5473 & 0.5576 & 0.3995 & 0.5444 & 0.6132 \\
    \bottomrule
    \end{tabular}
    }
    \label{tab:stra}
\end{table}

\subsubsection{Influence of Data Augmentation Strategy}
To study the effectiveness of our proposed sequence augmentation strategy, we test the performance on AOL with different combinations of strategies. The results are shown in Table~\ref{tab:stra}. ``\ttt{None}'' means that we use the original BERT parameters for document ranking without our proposed sequence optimization stage. We denote the term mask strategy as ``\ttt{TM}'', query/document deletion as ``\ttt{QDD}'', and behavior reordering as ``\ttt{BR}''. Note that the reordering strategy can only apply to sequences with more than two query-document pairs, thus cannot work independently.

First, compared with no sequence optimization stage, optimizing sequence representation with any combination of our proposed strategies is helpful. This clearly demonstrates that our proposed method is effective in building a more robust representation. Second, the term mask works best and this single strategy can improve around 2.5\% in MAP. This implies that learning user behavior sequences with similar queries and documents are very useful for document ranking. Finally, it is interesting to see that combining term mask and behavior reordering strategy (\ie, ``\ttt{TM + BR}'') leads to a performance degradation compared with only using the term mask strategy. After checking the sequence representation optimization process, we find that the contrastive learning loss in this case is very low and the prediction accuracy is very high, which indicates that this combination is easy to overfit and cannot learn a good sequence representation. 

\begin{table}[t!]
    \centering
    \small
    \caption{Performance of \ttt{COCA} on the AOL dataset with different hyperparameters.}
    \setlength{\tabcolsep}{1.5mm}{
    \begin{tabular}{cccccccc}
    \toprule
        & & CE ($\downarrow$) & Acc. & MAP & MRR & NDCG@1 & NDCG@10 \\
    \midrule
        \multirow{5}{*}{\rotatebox[origin=c]{90}{Temperature $\tau$}} 
        & 0.05 & \textbf{0.5662} & 79.62 & 0.5417 & 0.5521 & 0.3947 & 0.6078 \\
        & 0.1 & 0.5823 & 83.56 & \textbf{0.5500} & \textbf{0.5601} & \textbf{0.4024} & \textbf{0.6160} \\
        & 0.3 & 1.6240 & \textbf{84.03} & 0.5451 & 0.5552 & 0.3972 & 0.6116 \\
        & 0.5 & 4.3226 & 69.85 & 0.5433 & 0.5536 & 0.3950 & 0.6031 \\
        & 1.0 & 5.2148 & 62.33 & 0.5417 & 0.5522 & 0.3951 & 0.6073 \\
    \midrule
        \multirow{4}{*}{\rotatebox[origin=c]{90}{Batch Size}} 
        & 16 & 0.7289 & 81.14 & 0.5380 & 0.5482 & 0.3897 & 0.6044 \\
        & 32 & 0.7226 & 80.92 & 0.5447 & 0.5547 & 0.3972 & 0.6108 \\
        & 64 & 0.7210 & 81.20 & 0.5432 & 0.5534 & 0.3951 & 0.6089 \\
        & 128 & \textbf{0.5823} & \textbf{83.56} & \textbf{0.5500} & \textbf{0.5601} & \textbf{0.4024} & \textbf{0.6160} \\
    \bottomrule
    \end{tabular}
    }
    \label{tab:hyper}
\end{table}

\subsubsection{Performance with Different Hyperparameters}
As reported in recent work~\cite{SimCLR}, the temperature and batch size are two important hyperparameters in contrastive learning. To investigate the impact of them, we train our model with different settings and test their performance.
In addition to evaluating the performance of ranking, we also compute the loss value (cross-entropy, CE) and prediction accuracy in contrastive prediction. The results are shown in Table~\ref{tab:hyper}. 

Considering temperature, according to Equation (\ref{eq:cl}), a higher temperature will cause a higher loss, which are consistent with our results. However, a lower contrastive loss cannot always lead to a better performance. Indeed, $\tau=0.1$ is the best choice for the document ranking task. Therefore, it is important to select a proper temperature for contrastive learning. Similar observations are also reported in other recent studies~\cite{SimCLR,SimCSE}. 

As for batch size, we can see that contrastive learning benefits from larger batch sizes. 
According to a recent study~\cite{SimCLR}, larger batch sizes can provide more negative examples, so that the convergence can be facilitated. Due to our limited hardware resources, the largest batch size we can handle is 128. We speculate that a larger batch size can bring more improvements.
%
%
\begin{figure*}
    \centering
    \begin{subfigure}[b]{0.49\linewidth}
        \centering
        \includegraphics[width=\linewidth]{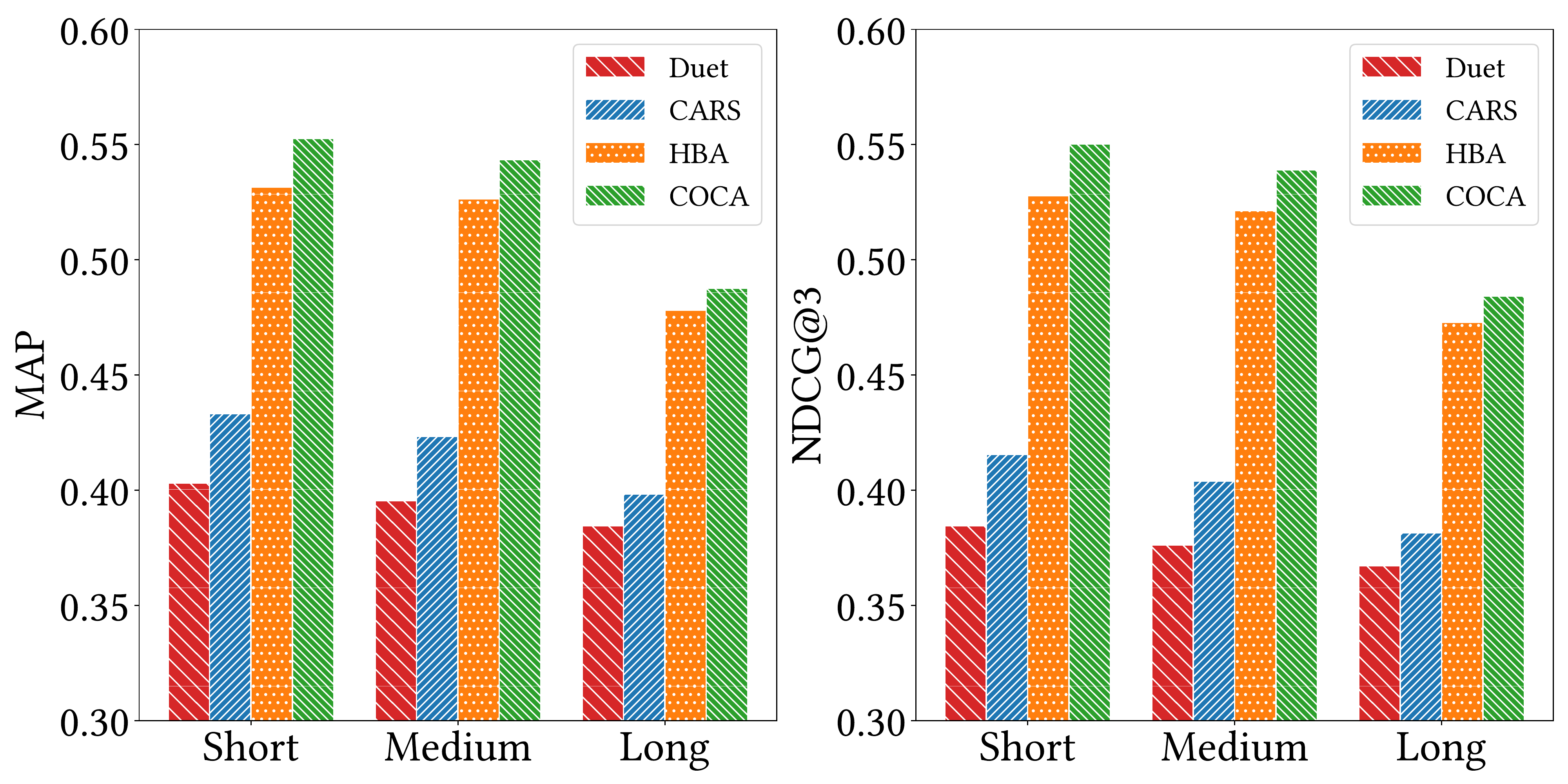}
    \end{subfigure}
    \begin{subfigure}[b]{0.49\linewidth}
        \centering
        \includegraphics[width=\linewidth]{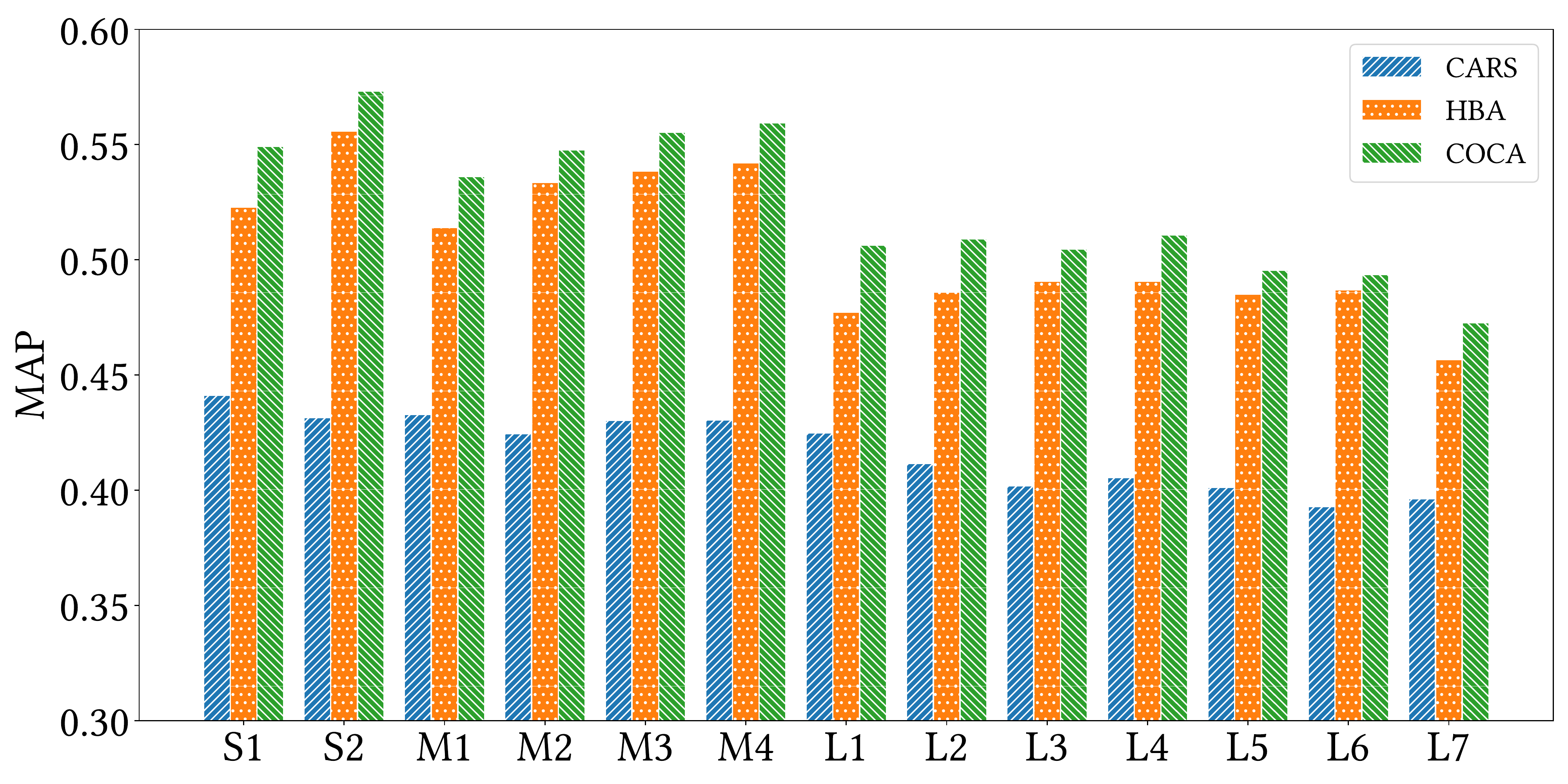}
    \end{subfigure}
    \caption{(Left) Performance on different lengths of sessions. (Right) Performance at different query positions in short (S1-S2), medium (M1-M4), and long sessions (L1-L7). The number after ``S'', ``M'', or ``L'' indicates the query index in the session.}%
    \label{fig:len}
\end{figure*}
\subsubsection{Performance on Sessions with Different Lengths}
To understand the impact of the session length on the final ranking performance, we categorize the sessions in the test set into three bins: 

(1) Short sessions (with 1-2 queries) - 77.13\% of the test set;

(2) Medium sessions (with 3-4 queries) - 18.19\% of the test set;

(3) Long sessions (with 5+ queries) - 4.69\% of the test set. \\
As we also consider sessions with only one query, the short sessions have a higher proportion than that provided in~\cite{CARS}.

We compare \ttt{COCA} with \ttt{Duet}, \ttt{CARS}, \ttt{HBA} on AOL dataset and show the results regarding MAP and NDCG@3 in the left side of Figure~\ref{fig:len}. First, it is evident that \ttt{COCA} outperforms all context-aware baseline methods on all three bins of sessions. This suggests \ttt{COCA}'s advantages in learning search context.
Second, we can see the ad-hoc ranking method \ttt{Duet} performs worse than other context-aware ranking methods. This demonstrates once again that modeling the historical user behavior is essential for improving the document ranking performance. 
Third, we can observe that \ttt{COCA} performs relatively worse in long sessions than in short sessions. We hypothesize that those longer sessions are intrinsically more difficult, and similar trend in baseline methods can support this. This can be due to the fact that a long session may contain more noise or exploratory search. This is also shown by a larger improvement in the short sessions from \ttt{COCA} to the ad-hoc baseline ranker \ttt{Duet} than that in the long sessions (37.10\% v.s. 26.83\% in terms of MAP). This result implies that it may be useful to model the immediate search context rather than the whole context.

\subsubsection{Effect of Modeling User Behavior Progression}
It is important to study how the modeled search context helps document ranking when a search session  progresses. We compare \ttt{COCA} with \ttt{CARS} and \ttt{HBA} at individual query positions in short (S), medium (M), and long (L) sessions. The results are reported in the right side of Figure~\ref{fig:len}. Due to the limited space, long sessions with more than seven queries are not presented.

It is noticeable that the ranking performance is improved steadily as a search session progresses, \ie, more search context becomes available for predicting the next click. Both \ttt{COCA} and \ttt{HBA} benefit from it, while \ttt{COCA} improves faster by better exploiting the context. In contrast, the performance of \ttt{CARS} is unstable. This implies that BERT-based methods are much more effective in modeling search context. One interesting finding is that, when the search sessions get longer (\eg, from L4 to L7), the gain of \ttt{COCA} diminishes. We attribute this to the more noisy nature of long sessions.
%
\begin{figure}
    \centering
    \includegraphics[width=\linewidth]{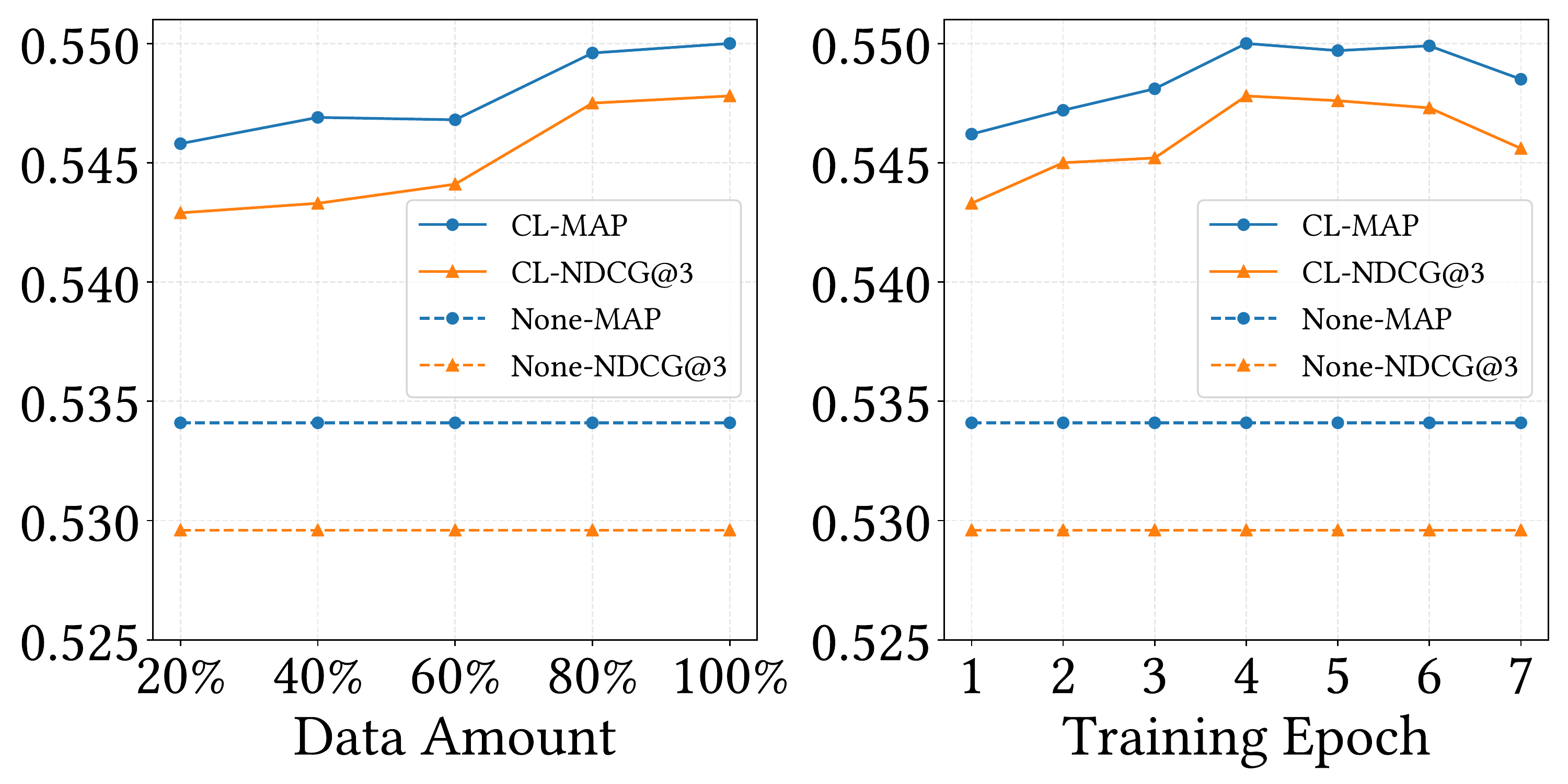}
    \caption{The performance with different training data amount and training epochs.}
    \label{fig:amount}
\end{figure}
\subsubsection{Influence of Amount of Training Data}
As reported by recent studies~\cite{SimCLR,SimCSE}, the amount of data for contrastive learning has a great impact on downstream task (\eg, document ranking in our case). We investigate such influence by training the model with different proportions of data and different epochs. As a comparison, we also illustrate the performance of \ttt{COCA} without sequence representation optimization stage (denoted as ``\ttt{None}''). 

We first reduce the number of training data used for contrastive learning\footnote{Note that all models are trained four epochs with different number of data.}. The results are shown in the left side of Figure~\ref{fig:amount}. It is clear that contrastive learning benefits from a larger amount of data. Surprisingly, our proposed sequence representation optimization stage can still work with only 20\% of training data. This demonstrates the potential and effectiveness of learning better sequence representation for context-aware document ranking. We also train \ttt{COCA} with different number of epochs in the sequence optimization stage. The performance on document ranking is shown in the right side of Figure~\ref{fig:amount}. The results suggest that the contrastive learning also benefits from larger training epochs. In our implementation, the data augmentation strategies are randomly selected in different epochs. Therefore, the sequence representation can be more fully learned. When training more than four epochs, the performance is stable without further improvement. Therefore, four epochs is the best choice in our experiments.



\section{Conclusion and Future Work}\label{sec:conclu}
In this work, we aimed at learning better representation of user behavior sequence for context-aware document ranking. A self-supervised task with contrastive learning objective is introduced for optimizing sequence representation before learning document ranking. To construct positive pairs in contrastive learning, we proposed three  data augmentation strategies at term, query/document, and user behavior level. These strategies can improve the generalization and robustness of sequence representation. The optimized sequence representation is used in document ranking task. We conducted comprehensive experiments on two large-scale search log datasets. The results clearly showed that our proposed method is very effective. In particular, our method with contrastive learning was shown to outperform the close competitor \ttt{HBA} without it.

This is the first attempt to utilize contrastive learning in IR and much remains to be explored. For example, it may be more appropriate to exploit recent history instead of the whole history. Query and document weighting in the history could also be a promising avenue.

\begin{acks}
Zhicheng Dou is the corresponding author. This work was supported by a Discovery grant of the Natural Science and Engineering Research Council of Canada, National Natural Science Foundation of China (No. 61872370 and No. 61832017),  Beijing Outstanding Young Scientist Program (No. BJJWZYJH012019100020098), Shandong Provincial Natural Science Foundation (No. ZR2019ZD06), and Intelligent Social Governance Platform, Major Innovation \& Planning Interdisciplinary Platform for the ``Double-First Class'' Initiative, Renmin University of China. We also wish to acknowledge the support provided and contribution made by Public Policy and Decision-making Research Lab of 
Renmin University of China.
\end{acks}

\clearpage
\balance

\end{document}